\documentclass[
 reprint,
superscriptaddress,
 amsmath,amssymb,prl
 aps
]{revtex4-2}

\usepackage{esint}
\usepackage{natbib}
\usepackage{amsmath,amssymb,amsfonts,graphicx,url,algorithm2e}
\usepackage[dvipsnames]{xcolor}
\usepackage{times}
\usepackage{dsfont}
\usepackage{mathtools}
\usepackage[utf8]{inputenc}
\usepackage[T1]{fontenc}
\usepackage{fancyhdr}

\usepackage{mathtools}
\usepackage{enumerate}
\graphicspath{{./},{./figures/}}

\usepackage{soul}

\newcommand{\Area}{{\mathcal A}}

\newcommand{\op}[1]{\operatorname{#1}}

\newcommand{\trace}{\op{Tr}}

\newcommand{\lam}{{l_c}}

\newcommand{\Lyap}{\mathcal{L}}
\newcommand{\nablax}{{\partial_x}}
\newcommand{\nablay}{{\partial_y}}

\newcommand{\magenta}{\color{black}}

\newcommand{\black}{\color{black}}

\usepackage{comment}
\usepackage{graphicx}\usepackage{epstopdf}
\begin{document}

\preprint{APS/123-QED}

\title{
  Energy harvesting from anisotropic fluctuations
}
\author{Olga Movilla Miangolarra}
\affiliation{%
Department of Mechanical and Aerospace Engineering, University of California, Irvine, CA 92697, USA
}
\author{Amirhossein Taghvaei}
\affiliation{%
Department of Mechanical and Aerospace Engineering, University of California, Irvine, CA 92697, USA
}%
\author{Rui Fu}%
\affiliation{%
Department of Mechanical and Aerospace Engineering, University of California, Irvine, CA 92697, USA
}
\author{Yongxin Chen}
\affiliation{%
School of Aerospace Engineering, Georgia Institute of Technology, Atlanta, GA 30332, \magenta{USA}
}%
\author{Tryphon T. Georgiou}
\affiliation{%
Department of Mechanical and Aerospace Engineering, University of California, Irvine, CA 92697, USA
}%

\date{\today}

\begin{abstract}
We consider a rudimentary model for a heat engine, known as the Brownian gyrator, that consists of an overdamped system with two degrees of freedom in an anisotropic temperature field.  Whereas the hallmark of the gyrator is a nonequilibrium steady-state curl-carrying probability current that can generate torque, we explore the coupling of this natural gyrating motion with a periodic actuation potential for the purpose of extracting work.  We show that path-lengths traversed in the manifold of thermodynamic states, measured in a suitable Riemannian metric, represent dissipative losses, while area integrals of a work-density quantify work being extracted. Thus, the maximal amount of work that can be extracted relates to an isoperimetric problem, trading off area against length of an encircling path.  We derive an isoperimetric inequality that provides a universal bound on the efficiency of all cyclic operating protocols, and a bound on how fast a closed path can be traversed before it becomes impossible to extract positive work.  The analysis presented provides guiding principles for building autonomous engines that extract work from anistropic fluctuations.
\end{abstract}

\maketitle

Harvesting energy is a principal characteristic of living organisms. Yet, relevant processes rarely conform to the setting of Carnot’s engine alternating contact between heat baths of different temperature. Instead, fluctuations and anisotropic chemical concentrations in conjunction with varying electrochemical potentials seem to provide the universal source of cellular energy \cite{battle2016broken,gnesotto2018broken}. The present work studies far-from-equilibrium transitions that are fueled by anisotropic thermal excitation, by adopting the frame of Stochastic Thermodynamics 
{{\cite{sekimoto2010stochastic,seifert2012stochastic,chen2019stochastic,Jarz2011ineq}}} and fluctuation theories \cite{Jarz1996eq,Cohen1995FT,Evans1994FT,Crooks1999FT,Hatano2001FT}.
Specifically, we study a minimal thermodynamic engine built around the concept of the Brownian gyrator \cite{BGyrator2007first}, a system that exhibits a characteristic non-equilibrium steady-state circulating current due to misalignment between the anisotropic temperature field and confining potential.

Previous work on the Brownian gyrator focused on the circulating current and torque generated at steady-state \cite{BGyrator2007first,BGyrator2013CilibertoExperim,BGyrator2013dotsenko,BGyrator2017electrical,BGyrator2017experimental,Imparato2017BG} and on optimal transitioning between states  \cite{baldassarri2020engineered}. In the present work, we take the next natural step to consider energetics of a cyclic operation. We utilize a controlled periodically time-varying potential to extract work from the anisotropy of the temperature field. To this end, we extend concepts of thermodynamic geometry \cite{Ruppeiner1995geom,Crooks2007length,Bradner2020geom} to regimes far-from-equilibrium. Specifically, we show that the length of a path in the two-dimensional Riemannian manifold of thermodynamic states represents dissipative losses, while the area integral of a work-density within a closed curve quantifies extracted work over the cycle. Thus, the problem to determine an optimal protocol reduces to an isoperimetric problem, where a path of a given length that encircles a maximal (weighted) area is sought. In this way, we quantify tradeoffs between efficiency and power that can be extracted.

{\em  Model and analysis:} 
We consider a two-dimensional overdamped Brownian particle in an anisotropic heat bath and subject to a time-varying potential $U(t,x,y)$, obeying the Langevin dynamics
\begin{subequations}	\label{eq:overdamped-Langevin}
\begin{align}
	d x_t &= - \gamma^{-1}\nablax U(t,x,y) d t + \sqrt{{2\gamma^{-1}k_BT_x}} d B^x_t,
\\
		d y_t &= - \gamma^{-1} \nablay U(t,x,y)d t + \sqrt{{2\gamma^{-1}k_BT_y}} d B^y_t,
\end{align}
\end{subequations}
where $\{B^x_t\}_{t\geq 0}$ and $\{B^y_t\}_{t\geq 0}$ are two independent standard Brownian motions, while $T_x$ and $T_y$ represent temperature along each of the two degrees of freedom $x$ and $y$, respectively. Throughout, $k_B$ denotes the Boltzmann constant, $\gamma$ a dissipation constant assumed identical in both directions, and $\nablax$ and $\nablay$ the partial derivatives with respect to $x$ and $y$, respectively. Without loss of generality, we assume $T_x >T_y$ and define $\Delta T :=T_x-T_y>0$. 
The probability distribution, that constitutes the state of the system, is denoted by $p(t,x,y)$ and satisfies the Fokker-Planck equation
    $
    \frac{\partial p}{\partial t}  + \nabla \cdot J = 0
    $
where 
\begin{equation*}\label{eq:flux}
    J =\left[\begin{matrix}J_x\\J_y\end{matrix}\right]= -\gamma^{-1}\left[\nabla U + \frac{1}{2} T \nabla \log(p)\right]p,
\end{equation*}
is the probability current, $\nabla$ is the gradient operator with respect to spatial coordinates, and 
\begin{equation*}
   T= \left[\begin{array}{cc}
         {2k_BT_x}&0  \\
         0& {2k_BT_y}
    \end{array}\right].
\end{equation*}

The system exchanges energy with the environment through work done by changes in the potential and through heat transfer with the two thermal baths. The total energy of the system is $E = \iint U p \, dx dy$, while the rate of work due to a change in the potential is given by
\begin{equation}\label{eq:work-p}
    \dot{W} = \iint \frac{\partial U}{\partial t} p\, dx dy.
\end{equation}
The heat uptake from the respective thermal baths is
\begin{align*}
    \dot{Q}_x &= \iint  J_x \nablax U \, dx dy
    = - \iint  U \nablax J_x \,dx dy ,\\
     \dot{Q}_y &= \iint    J_y\nablay U\,dx dy  = - \iint U \nablay J_y \,dx dy,
\end{align*}
resulting in the total heat uptake
\begin{equation}\label{eq:heat-p}
    \dot{Q} = \dot{Q}_x + \dot{Q}_y=- \iint  U \nabla \cdot J\,dx dy.
\end{equation}

Assume the potential is fixed and the system \eqref{eq:overdamped-Langevin} reaches a steady state. Stationarity only requires that $\nabla \cdot J=0$, implying zero total heat uptake. However, unless the detailed balance condition $J = 0$ holds, the steady-state is not an equilibrium distribution and the non-zero probability current mediates a steady-state heat transfer rate $\dot Q_x=-\dot Q_y\neq 0$ between the two thermal baths, which has been the subject of study of previous works~\cite{BGyrator2017electrical,Bgyrator2013ciliberto}.

In order to advance our analysis,
we henceforth assume a quadratic potential
\begin{equation*}
    U(t,x,y)=\frac{1}{2} \xi^\top K(t)\xi, \quad \mbox{where}\quad \xi=\left[\begin{array}{c}x\\ y\end{array}\right],
\end{equation*}
with $K(t)$ a symmetric $2\times 2$ matrix seen as a control variable. If the initial state is Gaussian, $N(0,\Sigma_0)$ (i.e., with mean $0$ and covariance $\Sigma_0$), then it remains Gaussian. Its mean remains $0$ while the
covariance $\Sigma(t)$ sastisfies the Lyapunov equation
\begin{equation}\label{eq:Lyapunov}
    \gamma \dot\Sigma(t)=-K(t)\Sigma(t)-\Sigma(t) K(t)+T.
\end{equation}
In terms of the state covariance and control, the energy is 
\[
E = \iint Updxdy=\frac{1}{2}\trace[K(t)\Sigma(t)],
\]
where $\trace [ \cdot ]$ denotes the trace. The rates of work input~\eqref{eq:work-p} and total heat input~\eqref{eq:heat-p} simplify to 
\begin{align*}
      \dot  W = \frac{1}{2}\trace[ \dot{K}(t) \Sigma(t)]\quad\mbox{and}\quad 
      \dot Q = \frac{1}{2}\trace[ {K}(t) \dot{\Sigma}(t)]. 
\end{align*}
Our goal is to design $K(t)$ so as to extract work by steering the covariance matrix $\Sigma(t)$ along a closed trajectory with $\{\Sigma(t);t\in[0,t_f],\Sigma(0)=\Sigma(t_f)\}$ in a cyclic manner.

To simplify our analysis we consider $\dot{\Sigma}(t)$ as our design parameter, instead of $K(t)$. We can do so since the unique $K(t)$ that satisfies~\eqref{eq:Lyapunov} is obtained in terms of $(\dot\Sigma(t),\Sigma(t))$ as
\begin{equation*}
    K(t) =
     \Lyap_{\Sigma(t)}[T - \gamma \dot{\Sigma}(t)],
\end{equation*}
where, for any positive definite matrix $A$, we define
\[
X \mapsto \Lyap_A[X]:=\int_0^\infty   e^{-\tau A} Xe^{-\tau A}d\tau.
\]
The heat rate, also expressed in terms of 
 $(\Sigma(t),\dot{\Sigma}(t))$, is
\begin{align*}
      \dot{Q}&=\frac{1}{2}\trace\bigg[\Lyap_{\Sigma(t)}[T]\dot\Sigma(t) \bigg]-\frac{\gamma}{2}\trace\bigg[\Lyap_{\Sigma(t)}[\dot \Sigma(t)]\dot\Sigma(t)\bigg].
\end{align*}
Integrating over $[0,t_f]$ we obtain that
    $
    Q = Q_{\rm qs} - Q_{\rm diss}
    $,
where
\begin{subequations}\label{eq:tot-heat}
\begin{align}
      Q_{\rm qs}&=\frac{1}{2}\int_0^{t_f} \trace\bigg[\Lyap_{\Sigma(t)}[T]\dot\Sigma(t) \bigg] d t,\\
      Q_{\rm diss} &= \frac{\gamma}{2} \int_0^{t_f}\trace\bigg[\Lyap_{\Sigma(t)}[\dot \Sigma(t)]\dot\Sigma(t)\bigg]dt.
\end{align}
\end{subequations}
These are integrals along the curve $\{\Sigma(t)\mid t\in[0,t_f]\}$. Note that the first one is independent of the time parameter while the second term converges to zero as the speed in traversing the path converges to zero. Thus, the first term corresponds to the effective heat uptake in the quasi-static limit and the second corresponds to dissipation.
When integrating over a cycle, the  work output is precisely their difference,
\begin{equation*}\label{eq:workout}
    W_{\rm out} = Q_{\rm qs} - Q_{\rm diss}.
\end{equation*}
Moreover, we define the efficiency of the cycle as the ratio between the work output and the maximum amount of work that can be extracted in a quasi-static setting \cite{Bradner2020geom},\footnote{This differs from the classical notion of efficiency $W_{\rm out}/Q_h$, where $Q_h$ is the heat taken from the hot heat bath.}, i.e.,
\[
\eta = \frac{W_{\rm out}}{Q_{\rm qs}}.
\]

 \begin{figure}[t]     \includegraphics[width=0.32\textwidth
 ]{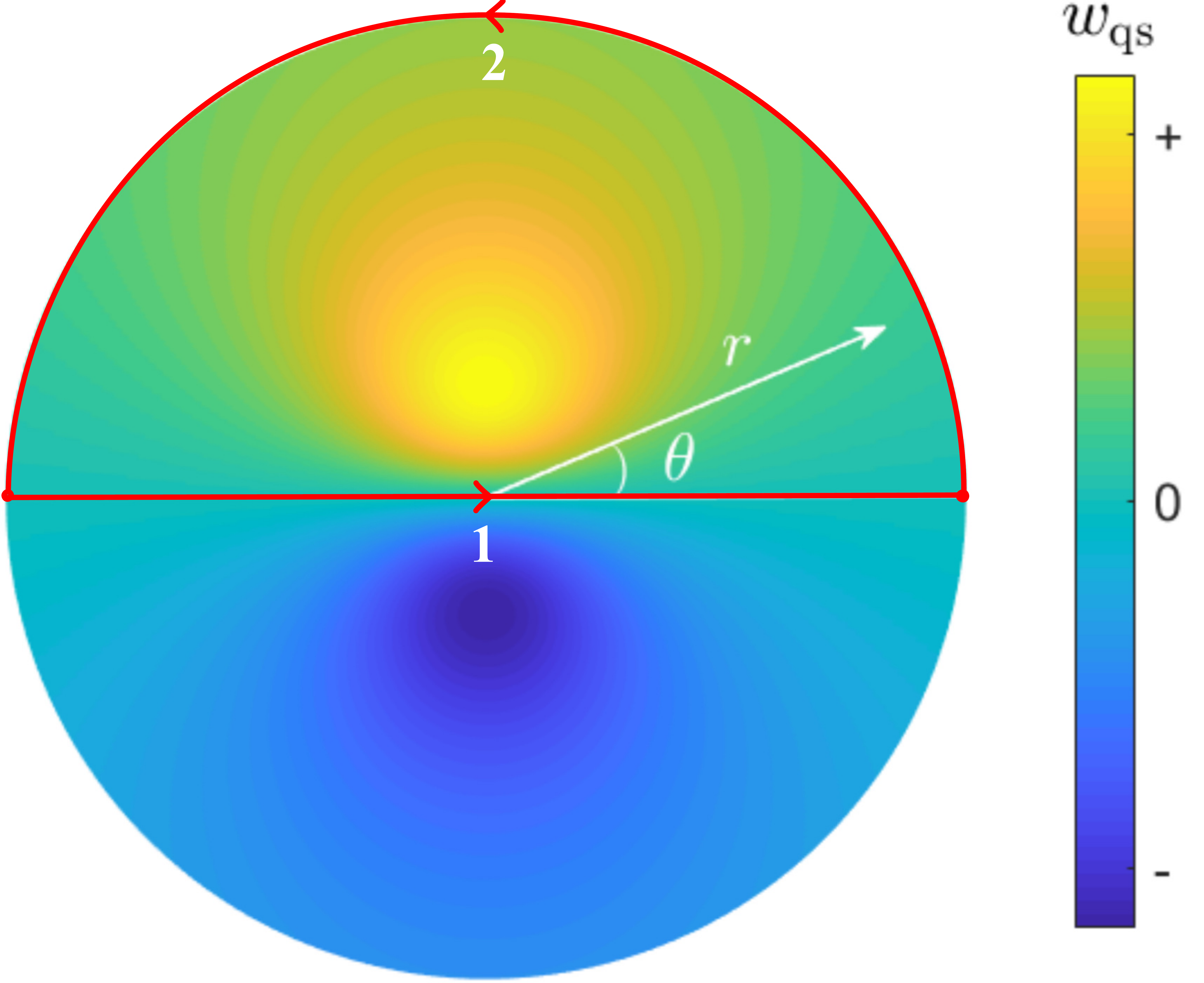}
        \caption{
        Work-density~\eqref{eq:w-qs} with values color coded, expressed in state-coordinates $(r,\theta)$ in \eqref{eq:Sigmartheta}. Area integrals over closed cycles represent quasi-static work. The red cycle encompasses the region of positive work-density within a given radius.}
        \label{fig:density}
\end{figure}

From this point on, we restrict the controlled degrees of freedom on the state manifold ($\Sigma$-space) to two by imposing that $\det(\Sigma(t))$ (or equivalently the entropy of the state) be constant. Under this restriction, the $2\times 2$ positive definite covariance (state) can be expressed in polar coordinates $(r,\theta)\in [0,\infty)\times [0,2\pi)$ as
\begin{equation}\label{eq:Sigmartheta}
    \Sigma(r,\theta)=R\Big(\hspace{-3pt}-\frac{\theta}{2}\Big)\sigma^2(r)R\Big(\frac{\theta}{2}\Big),
\end{equation}
where $R(\cdot)$ and $\sigma^2(\cdot)$ are orthogonal and diagonal matrices, respectively, given by
\begin{equation*}
    R(\vartheta)=\left[\begin{array}{cc}\cos(\vartheta)&\sin(\vartheta)\\-\sin(\vartheta)&\cos(\vartheta)\end{array}\right]~\mbox{and}~~
   \sigma^2(r) =\left[\begin{array}{cc}\lam^2e^r&0\\0&\lam^2e^{-r}\end{array}\hspace{-3pt}\right],
\end{equation*}
where $\lam=\sqrt[4]{\det(\Sigma(t))}$\label{page:charlength} is a (constant) {\em characteristic length} for the system. Therefore, the rate $\dot{\Sigma}$ can be expressed as the sum of two terms, one accounting for the rotation and the other for the expansion/contraction, that is,
\begin{align*}
    \dot{\Sigma} &= \frac{1}{2}R^\top(\sigma^2\Omega - \Omega \sigma^2)R \dot{\theta} +  R^\top\sigma^2\Xi R\dot{r},
\end{align*}
where
\begin{align*}
    \Omega=\begin{bmatrix}
    0 & 1 \\ -1 & 0
    \end{bmatrix}\quad\mbox{and}\quad 
    \Xi=\begin{bmatrix}
    1 & 0 \\ 0 & -1
    \end{bmatrix}.
\end{align*}
After substituting this expression for $\dot{\Sigma}$ into~\eqref{eq:tot-heat}, the quasi-static heat and dissipation can also be readily expressed in polar coordinates as follows,
\begin{subequations}
\begin{align}\label{eq:Q-qs-polar}
     {Q}_{\rm qs} &=\frac{k_B\Delta T}{2}\int_0^{t_f}\hspace*{-3pt} \left(\cos(\theta) \dot{r}-   \tanh(r) \sin(\theta)\dot{\theta} \right)dt,\\\label{eq:Q-diss-polar}
    {Q}_{\rm diss} &=\!\frac{\gamma { l_c^2}}{2}\! \int_0^{t_f} \! \left(\! \cosh(r)\dot{r}^2 \!+\! \sinh(r) \tanh(r) \dot{\theta}^2\!\right) dt.
\end{align}
\end{subequations} 

\begin{figure}[t]
    \centering
        \includegraphics[width=0.4\textwidth ,trim={1.85cm 1.74cm 1.85cm 0.5cm},clip]{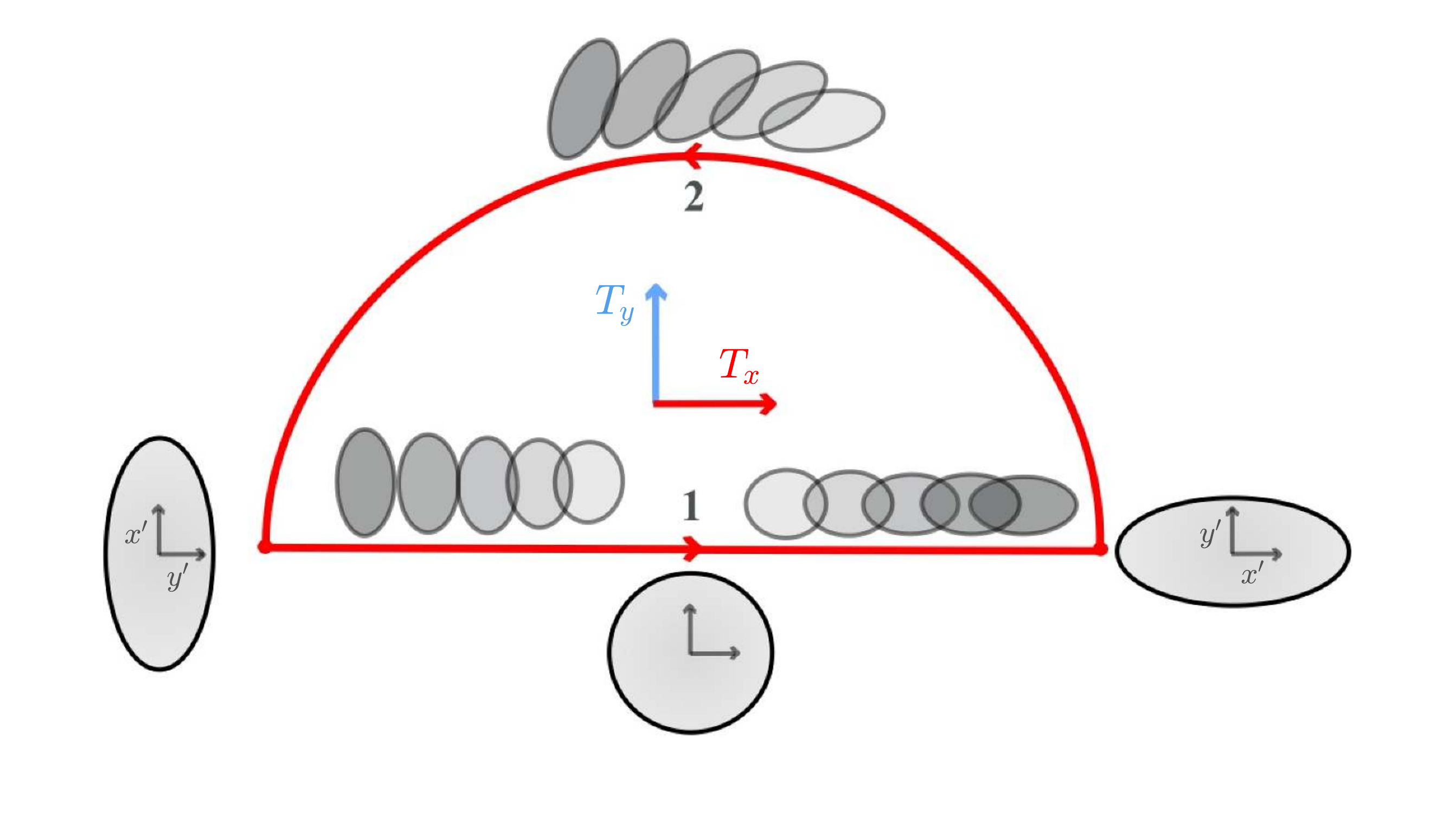}
        \caption{Cyclic protocol for semicircle path of radius $r_{\rm max}$ in Figure~\ref{fig:density}. The two phases represent: (1) expansion of the $\Sigma(t)$-ellipsoid along the $x$-axis with simultaneous compression along the $y$-axis, and (2) rotation to bring $\Sigma(t)$ to the starting value. The area of the ellipsoid remains constant during the cycle. When $T_x>T_y$, work is extracted during phase 1 and added during phase 2.}
        \label{fig:cycle}
\end{figure}

{\em Geometric interpretations:}
We now consider the integrals (\ref{eq:Q-qs-polar}-\ref{eq:Q-diss-polar}) over a cycle that encircles a domain $\mathcal D$, that is, over the boundary $\partial \mathcal D$ of $\mathcal D$. Using Stoke's theorem, $Q_{\rm qs}$ can be expressed as an area integral over $\mathcal D$,
\begin{align}\nonumber
    Q_{\rm qs} &=\frac{k_B\Delta T}{2} \oint_{\partial \mathcal D} \left(\cos(\theta)dr  - \tanh(r)\sin(\theta) d \theta \right) \\\label{eq:Qqscompute}
    &=  \pm \frac{k_B\Delta T}{2} \iint_{\mathcal D} \frac{\tanh^2(r)}{r}\sin(\theta) rd\theta d r,
\end{align}
where the sign is positive if the direction in traversing the cycle is counter clockwise (CCW), and negative otherwise. 
Thus,
\begin{equation}
    \label{eq:w-qs}
w_{\rm qs}(r,\theta)=\frac{k_B\Delta T}{2}\frac{\tanh^2(r)}{r}\sin(\theta),
\end{equation}
represents a quasi-static work density, which is depicted in Figure~\ref{fig:density}, and is positive on upper half plane and negative on the lower.  Any CCW cycle encircling a domain in the upper half plane results in positive work output. Likewise, a CW cycle in the lower half plane results in positive work output as well. The opposite is true when the flow is reversed. Below we always consider CCW-cycles.

The dissipation~\eqref{eq:Q-diss-polar} can be written as the (action) integral
\begin{equation*}
    Q_{\rm diss} =\frac{\gamma { l_c^2}}{2}  \int_0^{t_f} \| \dot{\alpha}(t) \|_g^2 dt,
\end{equation*}
where $\{\alpha(t) = (r(t),\theta(t))\mid t\in[0,t_f]\}$ traces $\partial \mathcal D$, and $\|\dot{\alpha}\|_g^2 := \dot{\alpha}^\top g\dot{\alpha}$ is the square norm of the velocity with respect to the Riemannian metric
 \begin{equation*}
      g=\left[\begin{array}{cc}
           \cosh(r)&0\\
         0&\sinh(r) \tanh(r)
     \end{array}\right].
 \end{equation*}
 By the Cauchy-Schwartz inequality, one obtains
 \begin{equation}\label{eq:Qdisscompute}
    Q_{\rm diss} \geq \frac{\gamma { l_c^2}}{2t_f}  \left(\int_0^{t_f} \| \dot{\alpha}(t) \|_g dt \right)^2,
\end{equation}
where equality holds when $\|\dot{\alpha}(t)\|_g$ remains constant. The integral in parentheses is the length of the closed curve $\{\alpha(t);t\in[0,t_f]\}$ in the metric $g$ \footnote{This equals the Wasserstein-2 length of the closed curve $\alpha(t)$.}.
From here on, we denote by $\mathcal M$ the Riemannian manifold of thermodynamic states equipped with the metric $g$.

\begin{figure}[t]
    \centering
    \includegraphics[width=0.32\textwidth,trim={0cm 0.25cm 0cm 0.3cm}, clip]{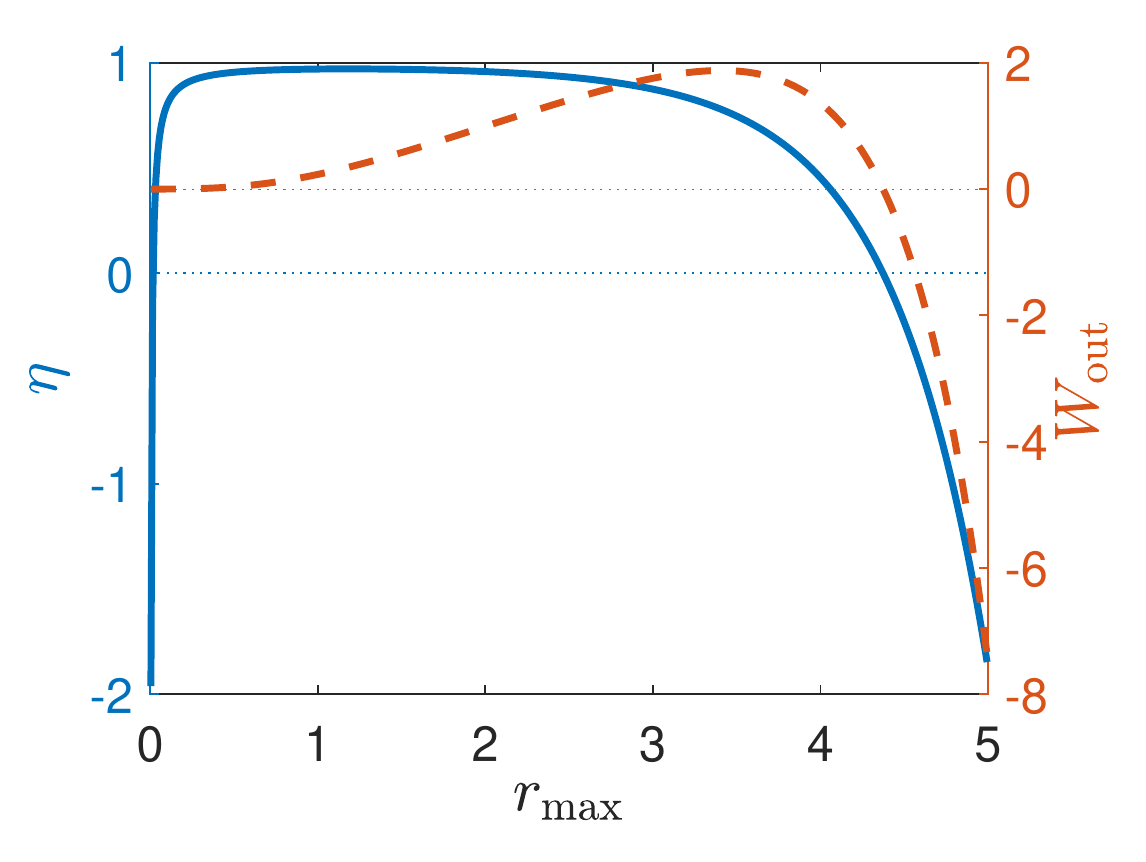}
    \caption{Efficiency and work output for the cycle depicted in \ref{fig:cycle} as a function of $r_{\rm max}$. }
    \label{fig:work-effic}
\end{figure}

The above results are exemplified in Figures~\ref{fig:cycle} and \ref{fig:work-effic}. Specifically, Figure~\ref{fig:cycle} displays $\Sigma(t)$-ellipsoids, relative to the principal axes of $T$, for the semicircle cycle (red) of Figure~\ref{fig:density}. Then, Figure~\ref{fig:work-effic} displays efficiency and work output for the same cycle, as a function of the radius of the semicircle, with the period of cycle fixed at $t_f=2\times10^{-3}$. Here, work is computed by subtracting the dissipation for constant velocity (RHS of \eqref{eq:Qdisscompute}) from the quasi-static work \eqref{eq:Qqscompute}.
Moreover, in Figure~\ref{fig:work-effic} we observe that an optimal value for $r_{\rm max}$ balances the two terms, the increase in area against increase in the perimeter, so as to maximize work output. This observation exposes an inherent isoperimetric problem that we discuss next.

Define the (weighted) area of $\mathcal D$  and its perimeter by
\begin{align*}
    \Area_f =\iint_{\mathcal D}f(r,\theta)\sqrt{\det(g)}d\theta d r,\quad 
    \ell = \oint_{\partial \mathcal D} \|\dot\alpha(t)\|_gdt,
\end{align*}
respectively, where 
\[
f(r,\theta)=\frac{\sin(\theta)\sinh(r)}{\cosh^2(r)},
\]
is a work-density relative to the Riemannian canonical $2$-form $\sqrt{\det(g)}d\theta dr$. 
 The area and the perimeter characterize the quasi-static heat 
 $Q_{\rm qs}$ and dissipation $Q_{\rm diss}$, as
  \begin{align*}
    Q_{\rm qs} = \frac{k_B\Delta T}{2} \Area_f,\quad Q_{\rm diss} = \frac{\gamma l_c^2}{2t_f} \ell^2,
\end{align*}
 and these determine the work output $W_{\rm out}$ and efficiency $\eta$, as
\begin{equation}
     W_{\rm out}=\frac{k_B\Delta T}{2}\big(\Area_f-\mu \ell^2\big)\quad\mbox{and}\quad \eta = 1-\mu\frac{\ell^2}{\Area_f},
     \label{eq:defefficiency}
\end{equation}
where
 $\mu=\frac{t_c}{t_f}$ is a dimensionless constant, with  $t_c = \frac{\gamma \lam^2}{k_B \Delta T}$ the {\em characteristic time} that a Brownian motion with intensity $\sqrt{\gamma^{-1}k_B\Delta T}$ needs to traverse a distance $\lam$ on average. 
 \begin{figure}[t]
     \centering
     \includegraphics[width=0.43\textwidth, trim={3cm 1.66cm 1.85cm 1.75cm}, clip]{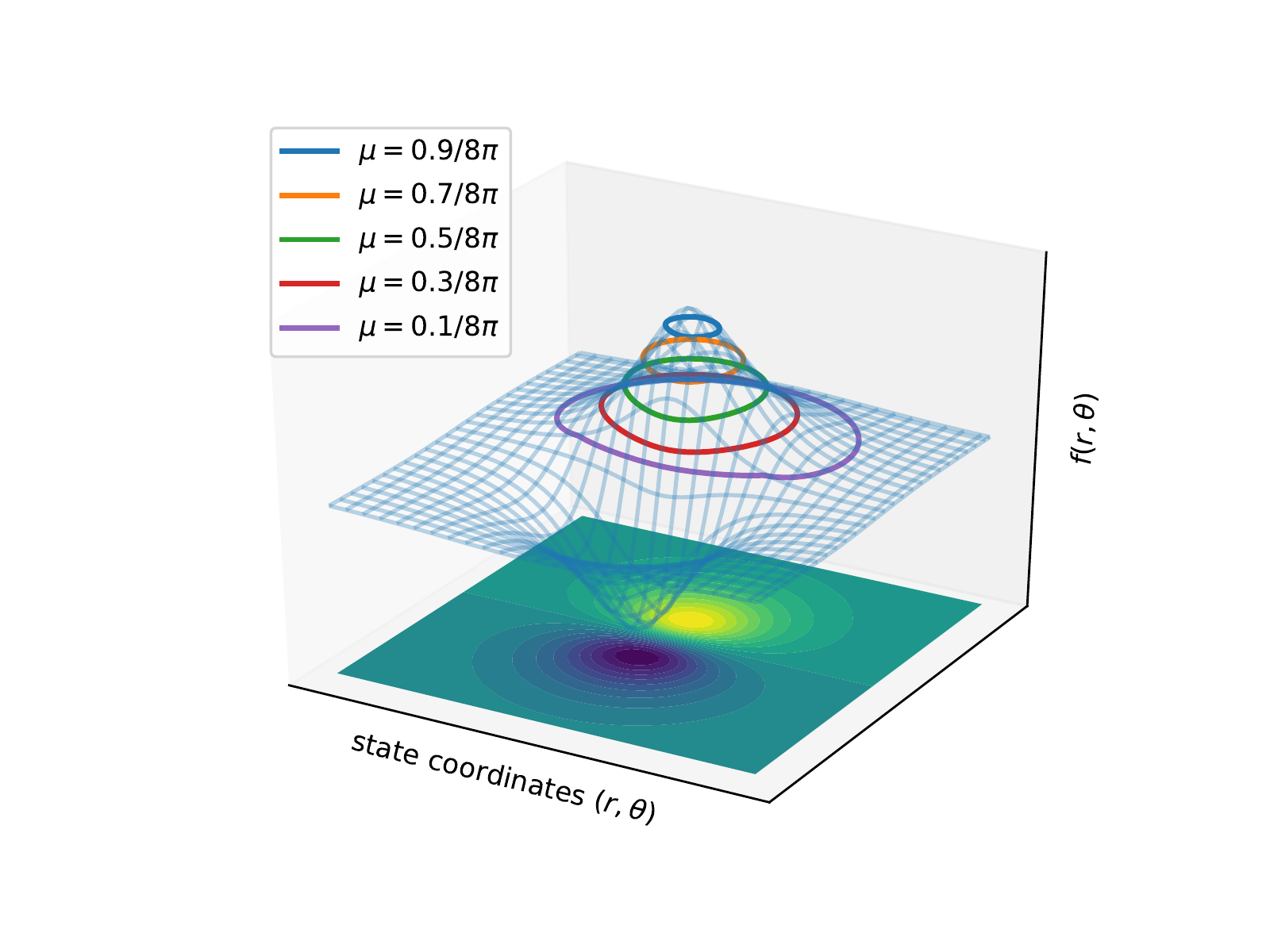}
     \caption{Optimal cycles, in polar coordinates $(r,\theta)$, that maximize work output for different values of $\mu$. The cycles are drawn on the $f$-density  surface and solve an isoperimetric problem. }
     \label{fig:max-work-amir}
 \end{figure}

We now consider maximizing work output over cycles on the manifold of thermodynamic states, i.e., to determine 
\begin{equation}\label{eq:maxwork}
    W^*(\mu):=\frac{k_B\Delta T}{2}\max_{\mathcal D} ~\{\Area_f - \mu \ell^2\},
\end{equation}
for different values of $\mu$. Maximization of $\Area_f-\mu\ell^2$ relates to the isoperimetric problem
\begin{equation}\label{eq:maxarea}
    \Area_f^*(\ell) :=\max_{\mathcal D}\{\Area_f \mid   \ell \mbox{ is specified }\},
\end{equation}
since
$\mu$ in \eqref{eq:maxwork} can be seen as a Lagrange multiplier for \eqref{eq:maxarea}.

We obtain a first-order condition that characterizes optimal cycles  through variational analysis. To this end, we parametrize the closed curve $\alpha(\cdot)$ tracing $\partial \mathcal D$ by the arclength $s$  and  let $ds$ and $du$ denote the unit differential along the curve and normal to the curve respectively. Under a perturbation $\alpha(s)\to \alpha(s) + \phi(s)\hat n(s)du$, where $\hat n(s)$ is the (outward) normal unit vector at $s$ and $\phi(\cdot)$ is an arbitrary scalar function, the perimeter is perturbed to $\int_{s=0}^\ell (1+\kappa(s) \phi(s)du)ds$, where $\kappa(\cdot)$ denotes the geodesic curvature~\citet{morgan1998riemannian}. Thus, the variation of $\ell^2$ is
$
\delta \ell^2 = 2\ell \int_{s=0}^\ell \kappa(s)\phi(s) dsdu
$. 
On the other hand, as the domain $\mathcal D$ is enlarged to $\mathcal D\cup\delta \mathcal D$,
\begin{align*}
\delta \Area_f &= \iint_{\delta\mathcal D}f(r,\theta)\sqrt{\det(g)}d\theta dr\\
&= \int_{s=0}^\ell  
f(r(s),\theta(s))\phi(s)dsdu.
\end{align*}
\black
Hence, the first-order optimally condition $\delta \mathcal A_f-\mu\delta \ell^2=0$ gives that the ratio of the geodesic curvature $\kappa$ over the density 
$f$
must be constant and equal to $1/(2\ell \mu)$ at each point of the curve that traces $\partial \mathcal D$.

Figure~\ref{fig:max-work-amir} displays several such optimal curves that have been obtained numerically using the first-order optimality condition. It is observed that as $\mu$ becomes small, and thus, the corresponding penalty on the length decreases,
the area that the optimal cycle encircles increases. On the other hand, as $\mu$ becomes large, the optimal cycle shrinks to the point $p_0=(r_0,\theta_0)=(\sinh^{-1}(1),\frac{\pi}{2})$, beyond which (i.e., for larger $\mu$) it is impossible to extract positive work. The point $p_0$ is where $f$ achieves its maximum. 

\begin{figure}[ht]
     \centering
     \includegraphics[width=0.4\textwidth,trim={0cm 0cm 0cm 0cm},clip]{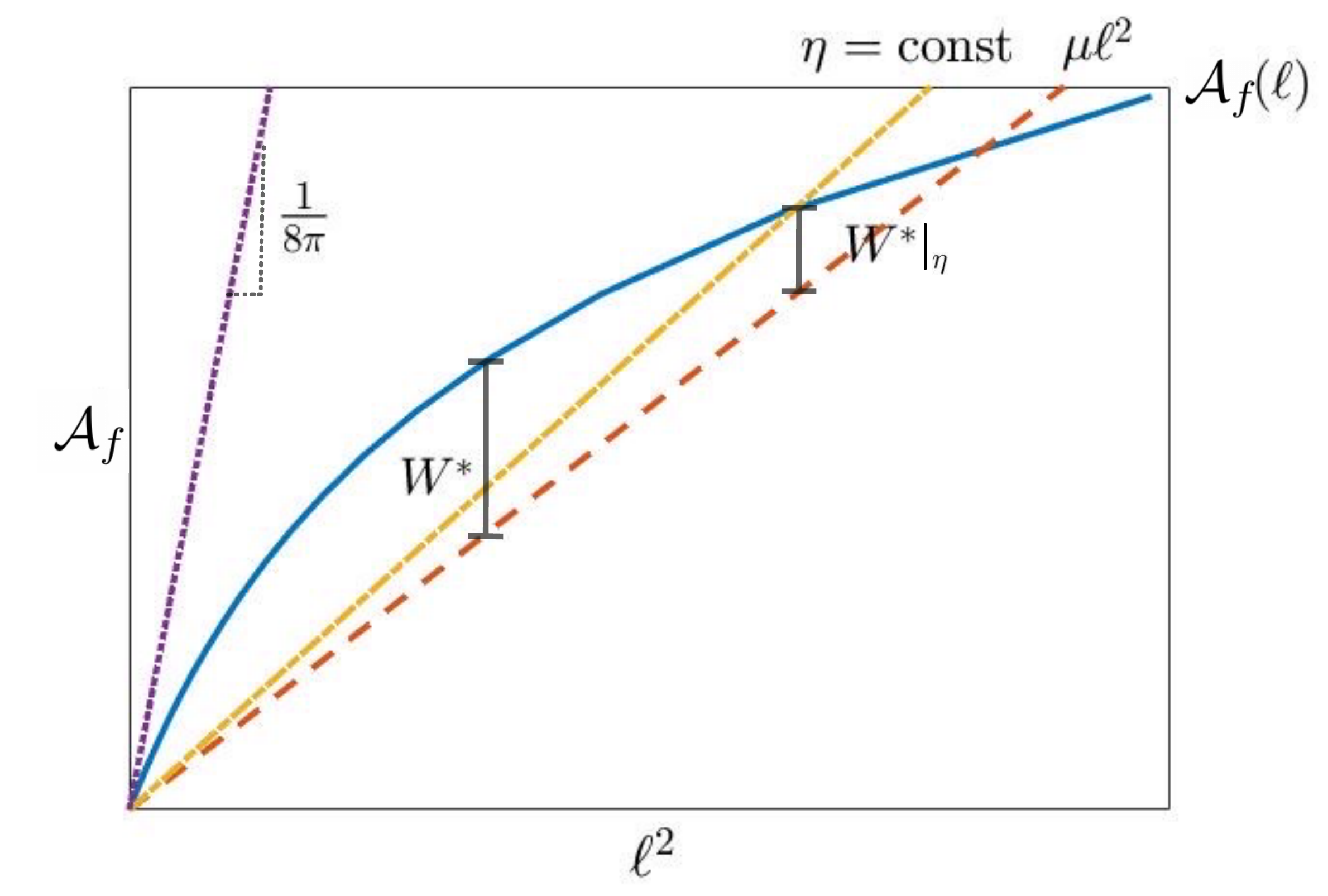}
     \caption{Maximum area $\mathcal A_f^*(\ell)$ after solving the isoperimetric problem \eqref{eq:maxarea}, shown with solid blue curve. }
     \label{fig:A-lsq}
 \end{figure}

The impossibility of extracting positive work for large values of $\mu$ points to an isoperimetric inequality that bounds the ratio between area and perimeter-squared, for all closed curves, with
\begin{equation}\label{eq:isoperimetric}
    \mu^*:=\sup_{\mathcal D}\bigg\{\frac{\Area_f}{\ell^2}\bigg\} <\infty,
\end{equation}
being the
isoperimetric constant.
In order to see this, we numerically evaluated the function $\mathcal{A}^*_f(\ell)$ in the isoperimetric problem \eqref{eq:maxarea} and reported the result in Figure~\ref{fig:A-lsq}. 
It can be seen from the figure that the ratio $\Area_f/\ell^2$ is maximized as $\ell\to 0$, which corresponds to vanishingly small cycles around $p_0$ in Figure~\ref{fig:max-work-amir}. For such cycles, the ratio can be analytically evaluated using local analysis, $
\Area_f/\ell^2\simeq f(p_0)/(4\pi)=1/(8\pi)$.
Thus we conjecture that $\mu^*= 1/(8\pi)$. Although the conjecture is not proven, we have established in the supplementary material that  $\mu^*\leq 1/4\pi$, and thus, finite.

The  isoperimetric inequality~\eqref{eq:isoperimetric} has two important implications. First, for $\mu \geq \mu^*$ (equivalently, $t_f\leq t_c/\mu^*$), it is impossible to extract positive work. Thus, $\frac{t_c}
{\mu^*}$ constitutes a threshold for the period of work producing cycles. 
Second, the efficiency is bounded by
\begin{equation*}
    \eta \leq 1-\frac{\mu}{\mu^*}= 1- \frac{1}{\mu^*}\frac{t_c}{t_f}.
\end{equation*}
The bound depends on physical parameters and the period, and turns negative when  positive work output is not possible.

The shape of $\Area^*_f(\ell)$ helps answer a variety of questions on optimizing protocols. 
Specifically, the maximal work output $W^*(\mu)$ in \eqref{eq:maxwork}
corresponds to the maximal vertical distance between  $\mathcal A_f^*(\ell)$ and the line $\mu \ell^2$, which takes place where $d\mathcal A_f^*(\ell)/d\ell^2=\mu$.
Also, it allows computing the maximal work for a given efficiency $\eta$. 
Operating points with efficiency $\eta$ provide work $W_{\rm out}=\eta \mathcal A_f$ and lie on the line $\mathcal A_f=\frac{\mu}{1-\eta}\ell^2$ shown (dash-dotted) in Figure~\ref{fig:A-lsq}. Therefore, the intersection of this line with the (blue) curve $\mathcal A_f^*(\ell)$ in Figure~\ref{fig:A-lsq} gives the sought optimal operating point for a given efficiency.

In the above, we tacitly assumed that the curve $\Area_f^*(\ell)$  intersects any line $\mu\ell^2$, for $\mu < \mu^*$, and that it eventually stays below the line, in that  $\lim_{\ell \to \infty}\Area_f/\ell^2 = 0$. We show that this is indeed true by proving the bound
\begin{equation}\label{eq:bound-Wout}
    \Area_f  -\mu \ell^2\leq \frac{1}{4\mu}
\end{equation}
for all $\mu>0$. 
This bound is established through a completion of squares argument in the supplementary material. Taking $\mu = \frac{1}{2\ell}$ in~\eqref{eq:bound-Wout}, we have that $\Area_f \leq \ell$, concluding that $\lim_{\ell \to \infty}\Area_f/\ell^2 = 0$. 
Another consequence of~\eqref{eq:bound-Wout} is that the power output is bounded as well, since
\begin{equation*}
    \text{power} = \frac{W_{\rm out}}{t_f} = \frac{k_B \Delta T}{2t_f}(\Area_f - \mu \ell^2) \leq \frac{k_B\Delta T}{8t_c}.
\end{equation*}
It is important to note that this bound on power is independent of the period $t_f$, and only depends on the ratio between the energy $k_B \Delta T$ and the characteristic time $t_c$.

We conclude with two directions for future work. The first pertains to the curvature of the thermodynamic manifold. It is known that a stronger isoperimetric inequality $\ell^2 \geq  \Area_f/\mu^* - G_f \Area_f^2$, with $G_f<0$, holds for spaces with everywhere negative Gaussian curvature~\cite{morgan1998riemannian}, \cite[page 1206]{osserman1978isoperimetric}. The concave shape of $\Area^*_f(\ell)$ suggests that a similarly strong inequality holds for $\mathcal M$, though at present, a proof is lacking.

A second direction pertains to the stability of optimal periodic protocols $K(t)$ that induce a nominal $\Sigma(t)$ via  \eqref{eq:Lyapunov}. Stability is the property of the state converging to the nominal cycle after any small perturbation, e.g.,  $\Sigma(0)\to \Sigma(0) + \Delta(0)$. From there on, the perturbation from the nominal cycle obeys
\[
\gamma \dot \Delta(t)=-K(t)\Delta(t)-\Delta(t)K(t).
\]
It can be shown that $\Delta(t)\to 0$  if the integral of the smallest eigenvalue of $K(t)$ over a period is positive; this is a standard argument and relies on showing that, under the eigenvalue condition,  $V(t):=\trace[\Delta(t)^2]$ decreases with time (Lyapunov function). We numerically verified that the optimal curves shown in Figure~\ref{fig:max-work-amir} satisfy the stated stability condition.  However, providing a theoretical guarantee for the stability of all optimal curves remains open and the subject of ongoing work. 

\bibliography{apssamp}\begin{widetext}
\newpage

\noindent{\sc Supplemental material}

\appendix
\section{Bounding $
\mu^*$}
Isoperimetric inequalities bound the area that can be encircled by closed curves of a given length and are inherently related to the Gaussian curvature of the space. In our case, by Gauss' celebrated theorema egregium \cite[page 23]{morgan1998riemannian}, the Gaussian curvature of the Riemannian manifold $\mathcal M$ can be computed, and it is
\begin{align}\nonumber
G(r,\theta)&=\frac{1}{\cosh^3(r)}.
\end{align}
Therefore, $\mathcal M$ is positively curved. However, the curvature decreases radially to $0$. For such manifolds, where in addition $g$ is rotationally symmetric, the following isoperimetric inequality holds
\cite[Page 113]{morgan1998riemannian},
\begin{equation}\label{eq:iso}
\ell^2 \hspace*{-2pt}\geq 4\pi\times {\Area} -2\int_0^{\Area}\bar G(\tau)d\tau
\end{equation}
where 
\[
\Area=\iint_{\mathcal D}
    \sqrt{\det(g)}drd\theta
\]
is the area of $\mathcal D$ with respect to the canonical $2$-form of $\mathcal M$, and
$\bar G(\tau)$ is the area integral of the Gaussian curvature over a circle centered at the origin with area $\tau$. This circle has radius $r(\tau)=\cosh^{-1}(1+\frac{\tau}{2\pi})$. Therefore,
\begin{align*}
   \bar G(\tau)&=\int_0^{r(\tau)}\int_0^{2\pi}\frac{\sinh(r)}{\cosh(r)^3}d\theta dr=\frac{\pi\tau}{2\pi + \tau}.
\end{align*}
Using this result in \eqref{eq:iso},
\begin{align*}
\ell^2&\geq 4\pi\Area- 4\pi^2\bigg(\frac{\Area}{2\pi}-\log\big(1+\frac{\Area}{2\pi}\big)\bigg)
\geq 2\pi \Area.
\end{align*}
Since  $   \Area_f\leq  \max_{(r,\theta)\in\mathcal D} ~f(r,\theta) \times \Area=\frac12 \Area
$, we conclude that
\[\mu^*\leq \frac{1}{4\pi}.
\]

    This bound is not tight due to the fact that $f$ is not rotationally symmetric (as opposed to the curvature) and achieves its maximum at $(r_0,\theta_0)=(\sinh^{-1}(1),\frac{\pi}{2})$.

    \section{Bounding work output}
    
    Using the expressions \eqref{eq:Q-qs-polar} and \eqref{eq:Q-diss-polar}, for $Q_{\rm qs}$ and $Q_{\rm diss}$, the work output
    $W_{\rm out}=Q_{\rm qs}-Q_{\rm diss}$ becomes
    \begin{align*}
       W_{\rm out}&=\frac{k_B\Delta T}{2}\int^{t_f}_0\big\{\cos(\theta) \dot{r}-   \tanh(r) \sin(\theta)\dot{\theta} -t_c\cosh(r)\dot{r}^2 \!-\! t_c\sinh(r) \tanh(r) \dot{\theta}^2\big\}dt\\
       &=\frac{k_B\Delta T}{2}\int^{t_f}_0\bigg\{-\bigg(\sqrt{t_c\cosh (r)}\, \dot{r}-\frac{\cos (\theta)}{2\sqrt{t_c\cosh (r)}}\bigg)^2\hspace{-2pt}-\bigg(\sqrt{t_c\sinh (r)\tanh (r)}\,\dot\theta+\frac{\sin (\theta)}{2\sqrt{t_c\cosh (r)}}\bigg)^2\hspace{-2pt}+\frac{1}{4 t_c \cosh (r)}\bigg\}dt\\
       &\leq \frac{k_B\Delta T}{8 t_c}\int^{t_f}_0\frac{1}{ \cosh (r)}dt\leq \frac{k_B\Delta T}{8 \mu},
    \end{align*}
   where in the first step we have completed the squares, and for the last inequality we used the fact that $\cosh (r)\geq 1$.

   \end{widetext}

\end{document}